\documentclass[aps,prb,preprint,groupedaddress]{revtex4}
\begin{document}
\title{On the derivation of the Dirac equation}
\author{R.Sartor}
\email[]{R.Sartor@ulg.ac.be}
\affiliation{
University of Liege,
Sart-Tilman,
Institute of Physics B5,
B-4000 Liege 1,Belgium}
\date{\today}
\begin{abstract}
We point out that the anticommutation properties of the Dirac matrices
can be derived without squaring the Dirac hamiltonian, that is, without 
any explicit reference to the Klein-Gordon equation. We only require
the Dirac equation to admit two linearly independent plane wave solutions
with positive energy for all momenta. The necessity of negative energies as
well as the trace and determinant properties of the Dirac matrices
are also a direct consequence of this simple and minimal requirement.
\end{abstract}
\maketitle

\section{Introduction}
Many textbooks \cite{bjorken,messiah,roman,scadron,schiff,thaller} 
derive the Dirac equation 
for a free particle of mass $m$ following
the method used by Dirac himself in his 1928 paper \cite{dirac}. This
method involves two steps. First, one admits that the
wave function should be a multi-component object, as in the
non-relativistic theory of spin, and that its time
evolution is ruled by a partial differential equation of first order
in both the time and space derivatives. Working in a system of
units where $\hbar=c=1$, we have

\begin{equation}
i{\partial \Psi\over\partial t}=H_D\Psi,
\label{d}
\end{equation}
where the Dirac hamiltonian $H_D$ is defined by
\begin{eqnarray}
H_D&=&\mbox{\boldmath$\alpha$}\cdot(-i\nabla)+\beta m\nonumber\\
&=&\sum_{k=1}^3\alpha_k(-i{\partial\over\partial x^k})+\beta m.
\end{eqnarray}
In this equation, the $\alpha_k$'s 
and $\beta$ are constant hermitian matrices. In the second step, one 
'squares' Eq. (\ref{d}) by acting on both sides of it with the operator 
$i{\partial\over\partial t}$. This yields
\begin{eqnarray}
-{\partial^2 \Psi\over\partial t^2}
&=&H_D(i{\partial\over\partial t}\Psi)\nonumber\\
&=&H_D^2\Psi.
\end{eqnarray}
Then one requires $H_D^2$ to be identical to the operator $-\Delta+m^2$,
thereby ensuring that each of the components of $\Psi$ satisfies the
Klein-Gordon equation.This implies the anticommutation relations
\begin{eqnarray}
\alpha_i\alpha_j+\alpha_j\alpha_i&=&2\delta_{ij},\nonumber\\
\alpha_i\beta+\beta\alpha_i&=&0,\nonumber\\
\beta^2=1.
\label{rel}
\end{eqnarray}
Starting from these results, one usually proceeds 
by showing that such matrices indeed exist when $\Psi$
is a four-component object and then one 'finds' that Eq. (\ref{d}) 
admits both positive and negative energy plane wave solutions. So
the Dirac equation does not solve the 'problem of negative energies' 
which appears when studying the Klein-Gordon equation.

However, it is difficult to be immediately convinced that this fact 
is not a mere consequence of the requirement appearing in the second
step of the above derivation. Thus, to discard any doubts about
the necessity of negative energies, it would be more satisfactory
to avoid the squaring of the hamiltonian $H_D$. In this 
paper, we show that this is indeed feasible.
\section{Necessity of negative energies}
In order to implement the program outlined at the end of the previous 
section, we require Eq. (\ref{d}) to admit two linearly independent 
plane wave solutions of the form
\begin{equation}
\Psi({\bf x},t)=u({\bf p})e^{i({\bf p}\,\cdot\,{\bf x}-E_{\bf p}\,t)},
\label{pw}
\end{equation}
where $E_{\bf p}$ is the positive energy associated with a free
particle of momentum ${\bf p}$, that is,
\begin{equation}
E_{\bf p}=+({\bf p}^2+m^2)^{1/2}.
\label{pose}
\end{equation}
Since our aim is to describe spin $1/2$ particles such as 
electrons, this requirement is both natural and minimal. By 
inserting Eq. (\ref{pw}) into Eq. (\ref{d}), we obtain
\begin{equation}
E_{\bf p}\,u({\bf p})=h_D({\bf p})\,u({\bf p}),
\end{equation}
with
\begin{equation}
h_D({\bf p})=\mbox{\boldmath$\alpha$}\cdot{\bf p}+\beta m.
\end{equation}
Note that $h_D({\bf p})$ is a matrix of numbers
whereas $H_D$ is a matrix of differential operators. Since 
we obviously discard the solution $u({\bf p})=0$, we see, from
the above requirement, that $E_{\bf p}$ should
be a double root of the eigenvalue equation pertaining to 
the matrix $h_D({\bf p})$. Thus, if we introduce the characteristic
polynomial of $h_D({\bf p})$
\begin{equation}
P_n(E)=dtm[E-h_D({\bf p})],
\label{char}
\end{equation}
we should have
\begin{equation}
P_n(E_{\bf p})=0,
\label{p}
\end{equation}
and
\begin{equation}
P_n'(E_{\bf p})=0,
\label{pp}
\end{equation}
where $P_n'$ is the derivative of $P_n$ with respect to $E$. The index $n$
in these equations stands for the degree of $P_n(E)$
or, equivalently, for the number of components of the wave 
function $\Psi$.

Let us now try to satisfy Eqs. (\ref{p}) and (\ref{pp}) within a 
two-component theory ($n=2$). We have
\begin{equation}
P_n(E)\equiv P_2(E)=E^2+c_1({\bf p})E+c_0({\bf p}),
\end{equation}
where the coefficients $c_1$ and $c_0$ are polynomials homogeneous
in $m$ and the components $p_1$, $p_2$, $p_3$ of the 
momentum ${\bf p}$. Eq. (\ref{pp}) yields
\begin{equation}
2E_{\bf p}+c_1({\bf p})=0.
\end{equation}
It is not possible to satisfy this equation for all momenta
since the square root $E_{\bf p}$ cannot be expressed as a polynomial.
Thus, a two-component theory is immediately ruled out. So, let us try
a Dirac equation with three components. Now, we have
\begin{equation}
P_n(E)\equiv P_3(E)=E^3+c_2({\bf p})E^2+c_1({\bf p})E+c_0({\bf p}),
\end{equation}
where the coefficients $c_2$, $c_1$ and $c_0$ are again polynomials homogeneous
in $m$ and the components of the momentum ${\bf p}$. Eqs. (\ref{p})
and (\ref{pp}) yield
\begin{equation}
E_{\bf p}^3+c_2({\bf p})E_{\bf p}^2+c_1({\bf p})E_{\bf p}+c_0({\bf p})=0,
\end{equation}
\begin{equation}
3E_{\bf p}^2+2c_2({\bf p})E_{\bf p}+c_1({\bf p})=0.
\end{equation}
Again using the fact that $E_{\bf p}$ cannot be expressed as a polynomial, we
see that these equations imply
\begin{equation}
E_{\bf p}^2+c_1({\bf p})=0,
\label{p3}
\end{equation}
\begin{equation}
c_2({\bf p})E_{\bf p}^2+c_0({\bf p})=0,
\end{equation}
\begin{equation}
3E_{\bf p}^2+c_1({\bf p})=0,
\label{pp3}
\end{equation}
\begin{equation}
c_2({\bf p})=0.
\end{equation}
Eqs. (\ref{p3}) and (\ref{pp3}) lead to $E_{\bf p}=0$ for all momenta. This
is not possible and, as a consequence, a three-component Dirac theory is 
also ruled out. Finally, let us turn to a four-component 
theory. Now,
\begin{equation}
P_n(E)\equiv P_4(E)=E^4+c_3({\bf p})E^3+c_2({\bf p})E^2+c_1({\bf p})E
+c_0({\bf p}),
\label{pc4}
\end{equation}
where our notations are similar to those used above in the 
two- and three-component cases. Eqs. (\ref{p}) and (\ref{pp}) yield
\begin{equation}
E_{\bf p}^4+c_3({\bf p})E_{\bf p}^3+c_2({\bf p})E_{\bf p}^2
+c_1({\bf p})E_{\bf p}+c_0({\bf p})=0,
\end{equation}
\begin{equation}
4E_{\bf p}^3+3c_3({\bf p})E_{\bf p}^2+2c_2({\bf p})E_{\bf p}
+c_1({\bf p})=0.
\end{equation}
These equations imply
\begin{equation}
E_{\bf p}^4+c_2({\bf p})E_{\bf p}^2+c_0({\bf p})=0,
\label{p4a}
\end{equation}
\begin{equation}
c_3({\bf p})E_{\bf p}^2+c_1({\bf p})=0,
\label{p4b}
\end{equation}
\begin{equation}
2E_{\bf p}^2+c_2({\bf p})=0,
\label{pp4a}
\end{equation}
\begin{equation}
3c_3({\bf p})E_{\bf p}^2+c_1({\bf p})=0.
\label{pp4b}
\end{equation}
From Eq. (\ref{pp4a}), we obtain
\begin{equation}
c_2({\bf p})=-2E_{\bf p}^2.
\label{c2}
\end{equation}
Inserting this expression into Eq. (\ref{p4a}) yields
\begin{equation}
c_0({\bf p})=E_{\bf p}^4.
\label{c0}
\end{equation}
Finally, comparing Eqs. (\ref{p4b}) and (\ref{pp4b}) leads to 
\begin{equation}
c_1({\bf p})=0
\label{c1}
\end{equation}
and
\begin{equation}
c_3({\bf p})=0.
\label{c3}
\end{equation}
If we insert these results back into Eq. (\ref{pc4}), we see
that the eigenvalue equation for $h_D({\bf p})$ reads
\begin{equation}
(E-E_{\bf p})^2(E+E_{\bf p})^2=0.
\label{eve}
\end{equation}
This shows that the positive energy solutions to the Dirac equation
will always be accompanied by
solutions with negative energy. To prove that the approach adopted in this 
paper is self-contained, we still have to derive the anticommutation 
relations (\ref{rel}). This is performed in the next section.
\section{Derivation of the anticommutation relations}
We now show that Eqs. (\ref{c2}), (\ref{c0}), (\ref{c1}) 
and (\ref{c3}) do indeed
imply Eqs. (\ref{rel}). We remark that once we have replaced $E_{\bf p}$
by its expression (\ref{pose}), all of these equations require
some polynomial homogeneous in $m$ and the components of ${\bf p}$ to 
vanish identically, that is for all momenta. This is possible only if 
all the polynomial coefficients are zero. We shall
rely repeatedly on this remark in what follows.
  
From Eqs. (\ref{char}) and (\ref{pc4}), we obtain  
\begin{eqnarray}
c_3({\bf p})&=&-Tr(h_D({\bf p}))\nonumber\\
&=&\sum_{k=1}^3p_kTr(\alpha_k)+m Tr(\beta),
\end{eqnarray}
where the symbol $Tr$ denotes the trace. Thus, Eq.(\ref{c3}) implies
\begin{equation}
Tr(\alpha_1)=Tr(\alpha_2)=Tr(\alpha_3)=Tr(\beta)=0.
\label{trace}
\end{equation}
Eqs.(\ref{char}) and (\ref{pc4}) also yield
\begin{equation}
c_0({\bf p})=dtm(h_D({\bf p})).
\end{equation}
Inserting this expression into Eq.(\ref{c0}) and considering the 
terms in $p_1^4$, $p_2^4$, $p_3^4$ and $m^4$ leads to
\begin{equation}
dtm(\alpha_1)=dtm(\alpha_2)=dtm(\alpha_3)=dtm(\beta)=1.
\label{dtm}
\end{equation}

To make things simpler, it is convenient to work in a
representation where the matrix $\beta$ is
diagonal. Note that Eqs.(\ref{trace})
and (\ref{dtm}) are representation independent. Consider the terms
in $m^3$ in Eq.(\ref{c1}) and in $m^2$ in Eq.(\ref{c2}). They yield
\begin{equation}
\beta_{11}\beta_{22}\beta_{33}+
\beta_{11}\beta_{22}\beta_{44}+
\beta_{11}\beta_{33}\beta_{44}+
\beta_{22}\beta_{33}\beta_{44}=0
\label{prod3}
\end{equation}
and
\begin{equation}
\beta_{11}\beta_{22}+\beta_{11}\beta_{33}+\beta_{11}\beta_{44}+
\beta_{22}\beta_{33}+\beta_{22}\beta_{44}+\beta_{33}\beta_{44}
=-2,
\label{prod2}
\end{equation}
respectively. Combining these equations with 
\begin{equation}
dtm(\beta)=\beta_{11}\beta_{22}\beta_{33}\beta_{44}=1,
\end{equation}
(see Eq.(\ref{dtm})), we obtain
\begin{equation}
(1+\beta_{11})(1+\beta_{22})(1+\beta_{33})(1+\beta_{44})=0
\end{equation}
and
\begin{equation}
(1-\beta_{11})(1-\beta_{22})(1-\beta_{33})(1-\beta_{44})=0.
\end{equation}
These equations show that one of the eigenvalues of $\beta$ is
equal to $+1$ and another to $-1$. Let us assume that $\beta_{11}=+1$ 
and $\beta_{33}=-1$. Taking Eqs.(\ref{trace}) and (\ref{dtm})
into account, this implies 
$\beta_{44}=-\beta_{22}$ and $\beta_{22}=\pm 1$. We shall assume 
that $\beta_{22}=+1$ and thus $\beta_{44}=-1$. We do not have to consider
other choices for the
diagonal elements to be put equal to $+1$ or $-1$ since this would
correspond to a mere rearrangement of 
the lines and columns of $\beta$. Thus, we have
\begin{equation}
\beta=\left(
\begin{array}{cccc}
1&0&0&0\\
0&1&0&0\\
0&0&-1&0\\
0&0&0&-1
\end{array}
\right).
\label{bstruc}
\end{equation}
Obviously,
\begin{equation}
\beta^2=1,
\label{b2}
\end{equation}
and it is easy to show that this equation implies
\begin{equation}
\alpha_i^2=1\hspace{5mm}(i=1,2,3).
\label{a2}
\end{equation}
Indeed, let us just imagine that we perform, on all the Dirac matrices, a 
unitary transformation which brings $\alpha_1$, say, into diagonal form. We
expect that the matrix $\beta$ will no longer be diagonal but
Eq.(\ref{b2}) will remain true because it is representation 
independent. We now proceed for $\alpha_1$ as we did above for $\beta$, that
is, we concentate on the terms in $p_1^3$ in Eq.(\ref{c1}) and in  
$p_1^2$ in Eq.(\ref{c2}). This will lead us to $\alpha_i^2=1$ which is 
also representation independent. Proceeding in this way for
$\alpha_2$ and $\alpha_3$, we prove the
other identities in Eq.(\ref{a2}). This trick can be used each time
we establish a representation independent identity even if we arrived at
that identity within a particular representation. In what follows, we 
go back to the representation in which $\beta$ is given by Eq.(\ref{bstruc})
and we derive the structure of $\alpha_1$ in that representation. For the 
moment, we drop the index $1$ to simplify our notations. Thus, $\alpha$
stands for $\alpha_1$. Consider Eq. (\ref{c2}). The terms in $m p_1$
and in $p_1^2$ give
\begin{equation}
-\alpha_{11}-\alpha_{22}+\alpha_{33}+\alpha_{44}=0
\label{mp}
\end{equation}
and
\begin{eqnarray}
&&\alpha_{11}\alpha_{22}+\alpha_{11}\alpha_{33}+\alpha_{11}\alpha_{44}+
\alpha_{22}\alpha_{33}+\alpha_{22}\alpha_{44}+\alpha_{33}\alpha_{44}
\nonumber\\
&&-|\alpha_{12}|^2-|\alpha_{13}|^2-|\alpha_{14}|^2-
|\alpha_{23}|^2-|\alpha_{24}|^2-|\alpha_{34}|^2
=-2,
\label{p2}
\end{eqnarray}
respectively. Consider now the terms in $m^2p_1^2$ in Eq.(\ref{c0}),
they give
\begin{eqnarray}
&&\alpha_{11}\alpha_{22}-\alpha_{11}\alpha_{33}-\alpha_{11}\alpha_{44}-
\alpha_{22}\alpha_{33}-\alpha_{22}\alpha_{44}+\alpha_{33}\alpha_{44}
\nonumber\\
&&-|\alpha_{12}|^2+|\alpha_{13}|^2+|\alpha_{14}|^2+
|\alpha_{23}|^2+|\alpha_{24}|^2-|\alpha_{34}|^2
=2.
\label{m2p2}
\end{eqnarray}
Adding Eqs.(\ref{p2}) and (\ref{m2p2}) yields
\begin{equation}
\alpha_{11}\alpha_{22}+\alpha_{33}\alpha_{44}
-|\alpha_{12}|^2-|\alpha_{34}|^2=0.
\label{p2m2p2}
\end{equation}
On the other hand, comparing Eq.(\ref{mp}) with Eq.(\ref{trace}) yields
\begin{equation}
\alpha_{22}=-\alpha_{11}
\end{equation}
and
\begin{equation}
\alpha_{44}=-\alpha_{33}.
\end{equation}
If we insert these results back into Eq.(\ref{p2m2p2}), we obtain
\begin{equation}
\alpha_{11}^2+\alpha_{33}^2
+|\alpha_{12}|^2+|\alpha_{34}|^2=0.
\end{equation}
Thus, we have
\begin{equation}
\alpha_{11}=\alpha_{22}=\alpha_{33}=\alpha_{44}
=\alpha_{12}=\alpha_{34}=0,
\end{equation}
and, restoring the index $1$, we see that the matrix $\alpha_1$
has the following structure:
\begin{equation}
\alpha_1=\left(
\begin{array}{cccc}
0&0&(\alpha_1)_{13}&(\alpha_1)_{14}\\
0&0&(\alpha_1)_{23}&(\alpha_1)_{24}\\
(\alpha_1)_{13}^*&(\alpha_1)_{23}^*&0&0\\
(\alpha_1)_{14}^*&(\alpha_1)_{24}^*&0&0
\end{array}
\right),
\label{astruc}
\end{equation}
where the non-vanishing elements are restricted by the condition
\begin{equation}
|(\alpha_1)_{13}|^2+|(\alpha_1)_{14}|^2+|(\alpha_1)_{23}|^2
+|(\alpha_1)_{24}|^2=2.
\end{equation}
Actually, this equation tells us nothing new since it can be
derived from Eq.(\ref{a2}).
An analogous proof shows that the matrices $\alpha_2$ and 
$\alpha_3$ have also this structure. We note that the
structure of the $\alpha_i$'s and of $\beta$ 
(see Eq.(\ref{bstruc})) imply  
\begin{equation}
\alpha_i\beta+\beta\alpha_i=0 \hspace{5mm}(i=1,2,3),
\label{ab}
\end{equation}
as can be checked simply by performing matrix multiplications. Since
these equations are representation independent, we conclude, using
the trick described after Eq.(\ref{a2}), that
we should also require 
\begin{equation}
\alpha_i\alpha_j+\alpha_j\alpha_i=0 \hspace{5mm}(i\neq j=1,2,3).
\label{aa}
\end{equation}
Eqs.(\ref{b2}), (\ref{a2}), (\ref{ab}) and (\ref{aa}) are
the anticommutation relations we were looking for. It is easy to
check that Eqs.(\ref{c2}), (\ref{c0}) and (\ref{c1}) do not
give rise to additional restrictions on the Dirac matrices. As an 
example, consider the terms in $p_1p_2$ in Eq.(\ref{c2}). They
impose
\begin{eqnarray}
&&(\alpha_1)_{13}^*(\alpha_2)_{13}+(\alpha_1)_{13}(\alpha_2)_{13}^*
 +(\alpha_1)_{14}^*(\alpha_2)_{14}+(\alpha_1)_{14}(\alpha_2)_{14}^*
\nonumber\\
&+&(\alpha_1)_{23}^*(\alpha_2)_{23}+(\alpha_1)_{23}(\alpha_2)_{23}^*
 +(\alpha_1)_{24}^*(\alpha_2)_{24}+(\alpha_1)_{24}(\alpha_2)_{24}^*
=0.
\end{eqnarray}
This equation can be written as
\begin{equation}
(\alpha_1\alpha_2+\alpha_2\alpha_1)_{11}+
(\alpha_1\alpha_2+\alpha_2\alpha_1)_{22}=0
\end{equation}
and, indeed, requires nothing new.
\section{Summary and comments}
In this paper, we have provided a method to derive the anticommutation 
properties of the Dirac matrices without relying on the squaring
of the Dirac hamiltonian. We have only required Eq.(\ref{d}) to admit 
two linearly independent  plane wave solutions with positive energy for all 
momenta. At an early stage in the derivation, we have seen that, despite this 
conservative requirement, it was not possible to rule out negative energy 
solutions, thereby establishing that these 
are not an artefact of
the standard derivation. It might also be interesting to note
that, within the method described in this 
paper, the trace and determinant properties of the Dirac matrices
appear in the course of the derivation and not as by-products
of the anticommutation relations. Finally, a few comments are appropriate
concerning our proof of the impossibility of a two-component Dirac 
equation. As is well known, such an equation appears in a space-time
with less than three space dimensions \cite{thaller} or in the study of 
massless fermions, where it is known as the Weyl 
equation \cite{bjorken,scadron}. This by no means 
contradicts our assertions. Indeed, no
impossibility arises in a two-component theory if one only requires
the Dirac equation to admit a single plane wave solution with positive 
energy for all momenta. In that case, only Eq.(\ref{p}) with $n=2$
has to be imposed and this yields 
\begin{equation}
E_{\bf p}^2+c_1({\bf p})E_{\bf p}+c_0({\bf p})=0.
\end{equation}
This equation implies
\begin{equation}
E_{\bf p}^2+c_0({\bf p})=0,
\end{equation}
and
\begin{equation}
c_1({\bf p})=0.
\end{equation}
From these equations, we see that the eigenvalue equation for 
$h_D({\bf p})$ reads
\begin{equation}
(E-E_{\bf p})(E+E_{\bf p})=0.
\end{equation}
Thus, we have a plane wave solution with positive
energy and another with negative energy. As a consequence, in a two-component 
theory, the 'twofold degeneracy' only corresponds to the existence
of antiparticles. The derivation of the properties of the 
Dirac matrices (actually, of the Pauli matrices, since we are now in a 
two-component theory) can be performed as in the previous section and
will not be repeated here.

%\bibliography{dirac}
% Create the reference section using BibTeX:
%\bibliography{basename of .bib file}
%\begin{references}
%\bibitem{bd}
%bjo
%\bibitem{m}
%m
%\end{references}
\end{document}